\documentclass[a4paper]{jpconf}
\usepackage[english]{babel}
\usepackage[utf8]{inputenc}
\usepackage[T1]{fontenc}
\usepackage{amsmath}
\usepackage{amssymb}
\usepackage[separate-uncertainty=true,
 range-units = single,
 range-phrase = --
]{siunitx}

\usepackage{graphicx}

\begin{document}
\title{CRPropa - A Toolbox for Cosmic Ray Simulations}

\author{R.~Alves~Batista$^{1}$, J.~Becker Tjus$^{2}$, A.~Dundovic$^{3}$, M.~Erdmann$^{4}$, C.~Heiter$^{4,5}$, K.-H.~Kampert$^{6}$, D.~Kuempel$^{4}$, L.~Merten$^{2}$, G.~Müller$^{4}$, G.~Sigl$^{3}$, A.~v.~Vliet$^{3,7,8}$, D.~Walz$^{4}$, T.~Winchen$^{9}$, M.~Wirtz$^{4}$}

\address{
	$^{1}$~Universidade de São Paulo, Instituto de Astronomia, Geofísica e Ciências Atmosféricas; Rua do Matão, 1226, 05508-090, São Paulo-SP, Brazil,
}
\address{
	$^{2}$~Ruhr-Universität Bochum, Theoretische Physik IV: Plasma-Astroteilchenphysik, Universitätsstrasse 150, 44801 Bochum, Germany,
}
\address{
	$^{3}$~Universität Hamburg, II Institut für Theoretische Physik, Luruper Chaussee 149, 22761 Hamburg, Germany,
}
\address{
	$^{4}$~RWTH Aachen University, III. Physikalisches Institut A Otto-Blumenthal-Str., 52056 Aachen, Germany,
}
\address{
	$^{5}$~Max Planck Institut für Radioastronomie, Auf dem Hügel 69, 53121 Bonn, Germany,
}
\address{
	$^{6}$~Bergische Universität Wuppertal, Department of Physics, Gaußstr. 20, 42097 Wuppertal, Germany,
}
\address{$^7$ Department of Astrophysics/IMAPP, Radboud University, P.O. Box 9010, 6500 GL Nijmegen, The Netherlands,}
\address{
	$^{8}$~DESY Zeuthen, Platanenallee 6, 15738 Zeuthen, Germany,
}
\address{
$^{9}$~Vrije Universiteit Brussel, Astrophysical Institute, Pleinlaan 2, 1050 Brussels, Belgium
}
\ead{tobias.winchen@rwth-aachen.de}

\begin{abstract}
The astrophysical interpretation of recent experimental observations of cosmic
rays relies increasingly on Monte Carlo simulations of cosmic ray propagation
and acceleration.  Depending on the energy range of interest, several different
propagation effects inside the Milky Way as well as in extragalactic space have
to be taken into account when interpreting the data.  With the CRPropa
framework we aim to provide a toolbox for according simulations. In recent
versions of CRPropa, the ballistic single particle propagation mode aiming
primarily at extragalactic cosmic rays has been complemented by a solver for
the differential transport equation to address propagation of galactic cosmic
rays.  Additionally, modules have been developed to address cosmic ray
acceleration and many improvements have been added for simulations of
electromagnetic secondaries.  In this contribution we will give an overview of
the CRPropa simulation framework with a focus on the latest improvements and
highlight selected features by example applications.
\end{abstract}

\section{Introduction}
Current and future detectors for ultra-high energy cosmic rays (UHECR),
neutrinos, and gamma-rays provide us with an increasing amount of
precision data. The Pierre Auger Observatory~\cite{PAO2015a} and Telescope
Array~\cite{Kawai2008} measured the spectrum~\cite{Fenu2017,Matthews2017} and composition~\cite{Bellido2017,Matthews2017} of
the UHECR flux at the highest energies, and also observed anisotropy in the
arrival direction of UHECR~\cite{Abbasi2014,Aab2017e}.
The IceCube neutrino detector measured a flux of extraterrestrial neutrinos~\cite{Aartsen2013a}, and, recently, observed a first coincident
event with a flaring gamma ray blazar~\cite{IceCube2018}.

However, understanding this wealth of data is no easy task. The place and mechanism of acceleration of cosmic rays  remains
unknown and the newest data challenges the so far prevailing models.
Accounting for propagation effects, the observations point to a surprisingly
hard spectral index and heavy composition at the accelerators~\cite{Aab2016l},
requiring at least extensions to the simplest models for the acceleration
process. Observation of mass composition at the ankle~\cite{Buitink2016}
indicate a surprisingly light composition that is at odds with simple models
for the transition between galactic and extragalactic cosmic rays, and even at
the lowest energies the observed particle spectra challenge simple models for
acceleration in supernovas~\cite{Marrocchesi2017}.
To develop -- and test -- the increasingly complex models required to
understand our data, detailed simulations of the propagation and acceleration
mechanisms are needed.

\section{The CRPropa Software}
The CRPropa software in its third major version~\cite{Batista2016} has been developed  as a versatile simulation tool
to efficiently develop predictions from astrophysical models. The code includes
a variety of components to set up simulations for the propagation of particles,
including models for energy losses in interactions and creation of secondary particles.
In particular, particles can be propagated through arbitrary three-dimensional
magnetic fields and be subjected to energy losses due to interactions in photon fields by 
pair-production, photo-meson production, photo-disintegration, due to nuclear decay, and due to the
adiabatic expansion of the universe. The full propagation can be performed
including the combination of both,  cosmological effects and three-dimensional
propagation. The simulation of the physical processes is limited to the highly relativistic
propagation regime where the rest-mass of the particles can be neglected. In
addition to modules subjecting the propagated particles to physical effects, a
large collection of tools to inject particles with desired properties into the
simulation respectively record the requested output are included.
Execution of the simulation utilizes shared-memory parallel architectures
using OpenMP. Parallelization on distributed-memory architectures can be
implemented by users e.g.\ using MPI.

The individual components are independent and can be combined by the user to a
customized simulation. Additional components not yet included in CRPropa can be
written by the users in Python or C++ and included in the simulations, usually
without requiring a new compilation of the CRPropa code. We strive to collect
the contributions of users to the code and either include them into CRPropa or
facilitate sharing of code amongst user by referencing them on the CRPropa
webpage.

Code development follows an open procedure. All changes to the code are tracked
using a public git repository. Discussions on bugs and future enhancements as
well as support questions are transparent in an online issue tracker.
Contributions to the code by users as well as any larger modification is
subjected to peer review before inclusion to not only minimize wrong
computations, but also enforce technical standards and code quality to ensure
the future maintainability of the project.  Guidelines for
contributions, and thus also for the review, are public and distributed
alongside the CRPropa code. In particular, unit-tests, inline documentation,
and also usage examples published on the CRPropa website are required.
CRPropa is licensed under the GPLv3~\cite{GPLv3}.

\section{Recent Developments and Example Applications}
\subsection{Galactic Cosmic Rays}
\begin{figure}[tb]
	\includegraphics[width=\textwidth]{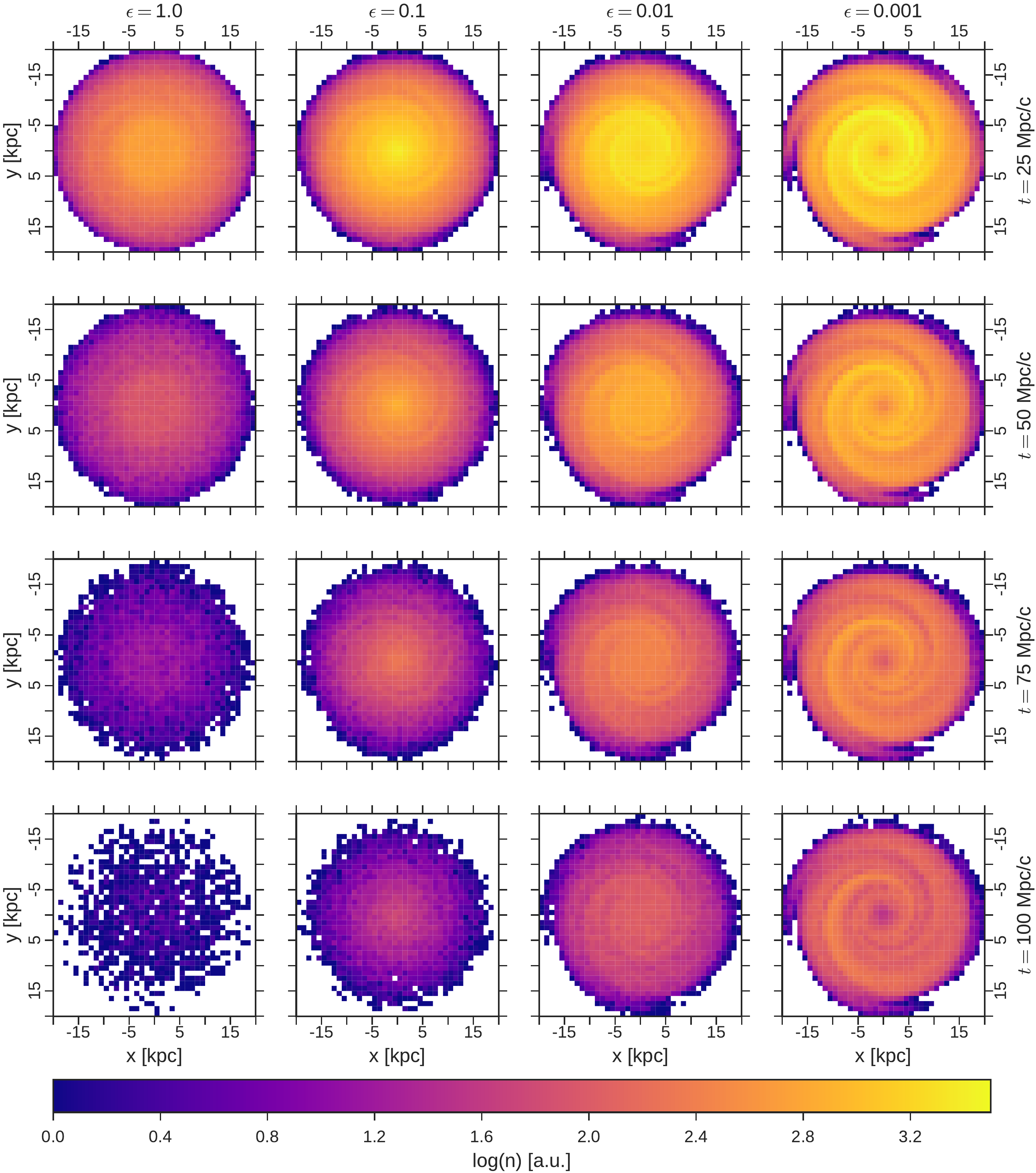}
	\caption{Time evolution of the density of cosmic rays in the Galactic plane resulting from homogeneous injection of cosmic rays in a JF 12 galactic magnetic field model for several choices of the diffusion parameter $\epsilon$ (From~\cite{Merten2017}).
	}
	\label{fig:galacticCosmicRayDensity}
\end{figure}

CRPropa has been recently extended to simulate propagation of galactic cosmic rays~\cite{Merten2017}.
For low energy cosmic rays the calculation of individual particle trajectories is not
feasible, as integrating the equation of motion becomes too computational intensive.
Instead, cosmic ray propagation is described stochastically via the modified Parker transport equation
\begin{equation}
	\frac{\partial n}{\partial t} + \vec{u}\cdot\nabla{n} = \nabla\cdot \left(\hat{\kappa}\nabla{n}\right) + \frac{1}{p^2}\frac{\partial}{\partial p}\left(p^2 \kappa_{pp} \frac{\partial n}{\partial p} \right) + \frac{1}{3}\left(\nabla \vec{u}\right)\frac{\partial n}{\partial \ln{p}} + S\left(\vec{x},p,t\right)
	\label{eq:ParkerTransport}
\end{equation}
which relates the particle density $n$  at point
$\vec{x}$, momentum $p$, and time $t$ of a system with the local spatial diffusion tensor
$\hat{\kappa}$, momentum diffusion coefficient $\kappa_{pp}$, advection speed
$\vec{u}$ and particle sources  $S\left(\vec{x},p,t\right)$. Additional terms
can be added to eq.~\ref{eq:ParkerTransport} to account for particle
interactions and decays.

The stochastic description can be reconciled with the single particle
description in CRPropa by treating the corresponding  stochastic
partial-differential equation instead of eq.~\ref{eq:ParkerTransport}. Here, individual `pseudo' particles make in
every step of the simulation  effective steps in space and momentum according
to the diffusion coefficients. In the limit of many particles the resulting
distributions of the particles are identical to the direct solution of
eq.~\ref{eq:ParkerTransport}, respectively the simulation of ensembles of
individual trajectories.

While this has the advantage that the propagation of ensembles of low energy
particles can be calculated very efficiently and interactions can be accounted
for at the same time with the regular CRPropa modules, additional assumptions
on the diffusion coefficients have to be made in the simulations.  As an
example application, in figure ~\ref{fig:galacticCosmicRayDensity} the time
evolution of the cosmic ray density in the Galactic disc is displayed for
several choices of the diffusion coefficient. In case of large 
ratios between parallel and perpendicular diffusion  $\kappa_\perp/\kappa_\parallel=\epsilon$, particles leak
quickly out of the galaxy; For small $\epsilon$ particles are bound to the
galactic field and follow the structure of the magnetic field models as expected.


\subsection{Particle Acceleration}
\begin{figure}[b]
	\centering
	\includegraphics[width=.6\textwidth]{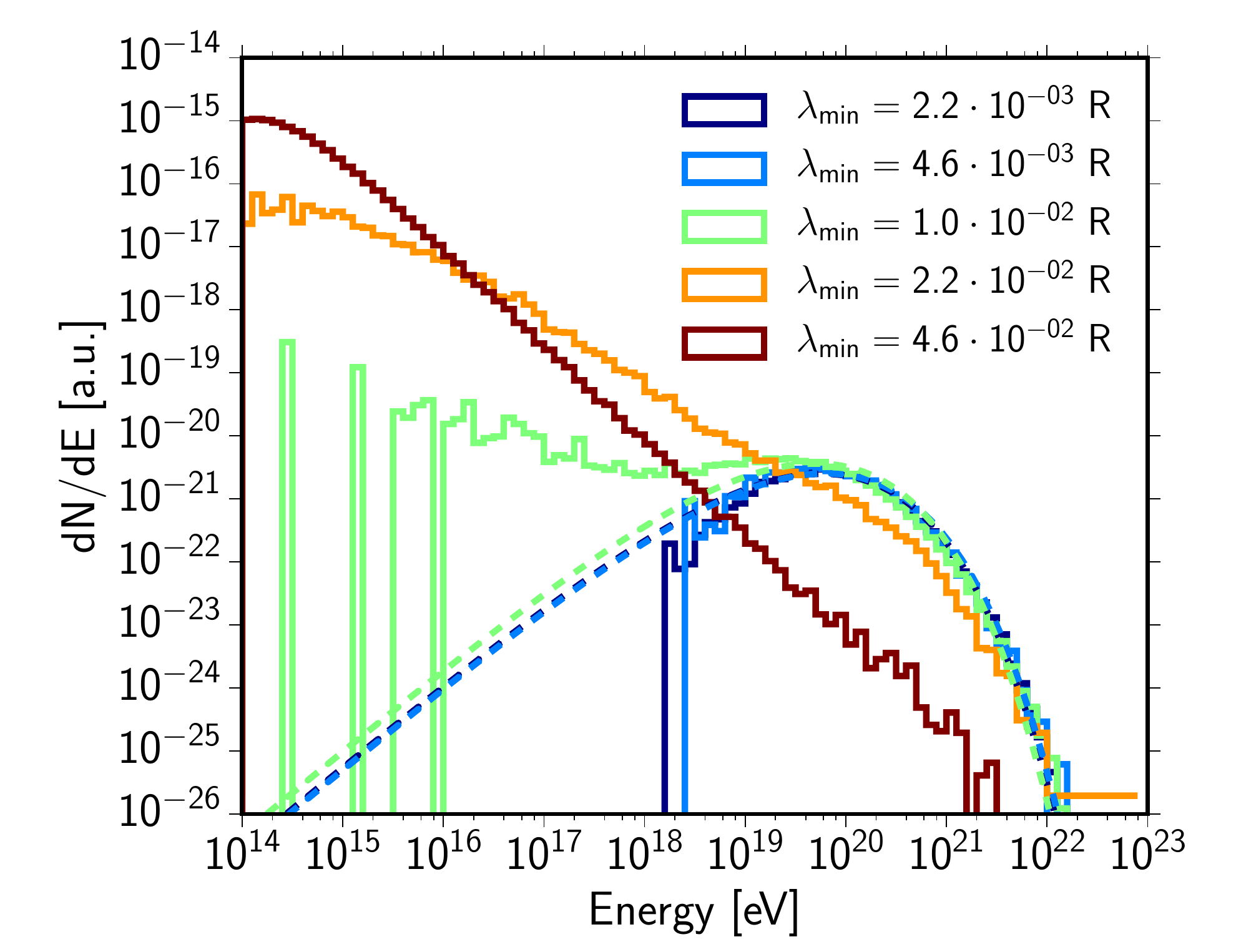}
	\caption{Acceleration spectra from second order Fermi acceleration with a
fixed minimum step length after injection in center of spherical
acceleration region with radius R~(From~\cite{Winchen2018}).}
	\label{fig:AccelerationSpectra}
\end{figure}
In the majority of models for cosmic ray sources, particles are believed to be accelerated in
shocked plasmas via the first-order Fermi mechanism~(e.g.~\cite{Kotera2011})
which predicts a power-law emission spectrum $dN/dE\propto E^{\gamma}$ with
$\gamma \leq -2$. However, evaluation of the data of the Pierre Auger
Observatory indicates~\cite{Aab2016l, Allard2008, Aloisio2014} harder
injection spectra that require new or at least modified acceleration
models~(e.g.~\cite{Unger2015,Winchen2018}).

To simulate the processes in the sources, we created modules to calculate
particle acceleration via scatter processes, e.g.\ by plasma waves. Input to
the simulation is a user defined step length corresponding to the frequency
of scatter events of the particles.  With appropriate choices of these parameters, the simulation of first-order
Fermi acceleration~\cite{Ghosh2017a} and second-order Fermi
acceleration is possible. Using this new feature it was demonstrated that
acceleration via the second-order Fermi mechanism after pre-acceleration,
geometrical effects of the acceleration region can modify the spectrum from a
power-law to a peaking distribution as shown in
figure~\ref{fig:AccelerationSpectra}. With the modified spectrum, the
observations of the Pierre Auger Observatory can be explained without implying
an unexpected high abundance of heavy elements at the sources~\cite{Winchen2018}.

\subsection{Photon Propagation and Production}
Photon propagation in CRPropa relied so far on external codes such as
DINT~\cite{Lee1998} or EleCa~\cite{Settimo2013}. Consequently, the simulations
were not easily extendable and full consistency between both components were
hard to achieve, as for example some photon models were not available in the
independent codes and effects of the local magnetic field were not easy to
account for.

To increase the precision of photon propagation as important secondary messenger,
new photon propagation modules have been created inside the CRPropa core structure
that allow a consistent simulations setup. For this, modules for all energy loss
process present in DINT and ELeCa,  have been implemented, and also additional
photon production channels are now considered ~\cite{Heiter2018}. In particular,
photon creation during photo-disintegration of nuclei, elastic scattering and
radiative decay have been included and the simulation accounts for deflections
in the local magnetic field. To also speed up the computation, which,
in some circumstances, can be otherwise prohibitively time consuming, a new
thinning procedure is currently under development.

\section{Conclusion}
The wealth of data in astroparticle physics provides new detailed insights in
the highest energetic processes of the universe. The CRPropa software framework
provides the necessary means to develop our understanding and compare the
predictions of the models with data.  The latest development efforts extend the
scope of the project from propagation of the highest energetic particles only
to include also lower energy and acceleration processes, thus enabling a
consistent view on high energy astroparticle physics.

\section*{Acknowledgements}
Tobias Winchen is supported by the German Research Foundation (DFG), grant WI 4946/1-1.
\section*{References}

\providecommand{\newblock}{}

\end{document}